\documentclass[prb,twocolumn,groupedaddress]{revtex4}
\usepackage{graphicx}
\begin{document}
\title{The favoured cluster structures of model glass formers}
\author{Jonathan P.~K.~Doye}
\email{jpkd1@cam.ac.uk}
\affiliation{University Chemical Laboratory, Lensfield Road, Cambridge CB2 1EW, United Kingdom}
\author{David J.~Wales}
\affiliation{University Chemical Laboratory, Lensfield Road, Cambridge CB2 1EW, United Kingdom}
\author{Fredrik H.~M.~Zetterling}
\affiliation{Department of Numerical Analysis, Royal Institute of Technology, 
SE-100 44 Stockholm, Sweden}
\author{Mikhail Dzugutov}
\affiliation{Department of Numerical Analysis, Royal Institute of Technology, 
SE-100 44 Stockholm, Sweden}
\date{\today}
\begin{abstract}
We examine the favoured cluster structures for two new interatomic potentials,
which both behave as monatomic model glass-formers in bulk.
We find that the oscillations in the potential lead to
global minima that are non-compact arrangements of linked 13-atom icosahedra.
We find that the structural properties of the clusters correlate with the glass-forming propensities 
of the potentials, and with the fragilities of the corresponding supercooled liquids. 
\end{abstract}

\maketitle

\section{Introduction}

There is still much to learn about the nature of the glass transition,\cite{Angell00}
and so it is an area of intense current interest. One of the increasingly prominent
avenues of research in this field is the computational study of model supercooled liquids 
and glasses,\cite{Debenedetti01} driven by the increase in computational power that has allowed larger systems 
to be studied over longer time scales. The advantage of the computational approach is that 
the configuration of the liquid or glass is directly available, potentially making it possible to 
understand the basis of the changes that occur as the glass transition is approached in terms of the 
structure and energy landscape. 
Indeed, this mode of research has led to an increasing number of new 
insights, for example, into the relationship between the energy landscape and the 
properties of supercooled liquids\cite{Sastry98,Sciortino99a} (in particular the fragility\cite{Sastry01}), 
the mechanisms of cooperative motion,\cite{Donati98}
the growing length scale of correlated motion as the temperature is decreased\cite{Benneman99} 
and whether a thermodynamic transition underlies the glass transition.\cite{Santen00}

One of the basic requirements of a model system to study 
supercooled liquids and glasses is that it will not crystallize 
on the relevant time scales. This constraint precludes the use of a monatomic system
interacting with simple pair potentials, such as the Lennard-Jones 
potential, because crystallization can occur relatively easily. 
Instead, the most commonly used model glass former is a binary Lennard-Jones 
system developed by Kob and Andersen,\cite{Kobmodel}
However, one drawback of using a binary system is that it is much harder to 
identify the structural changes that occur on approaching the
glass transition. Indeed the nature of the preferred local order in this system 
has not been systematically studied.  Furthermore, 
a crystalline ground state has recently been discovered for this system.\cite{Middleton01b}
In monatomic Lennard-Jones systems it is relatively 
easy to detect the presence of any close-packed crystallinity in a sample, but
incipient crystallization is much less obvious in a binary system with a complex
crystal structure, where the growth rate of any crystalline nuclei 
is expected to be slow. It is therefore unclear whether incipient 
crystallization could have contaminated any of the previous studies
of the supercooled liquid of the binary Lennard-Jones system. 
In one case, at least,
a low-energy configuration has been generated that shows signs of demixing 
and crystalline layering.\cite{Middleton01}

\begin{figure}
\begin{center}
\includegraphics[width=8.5cm]{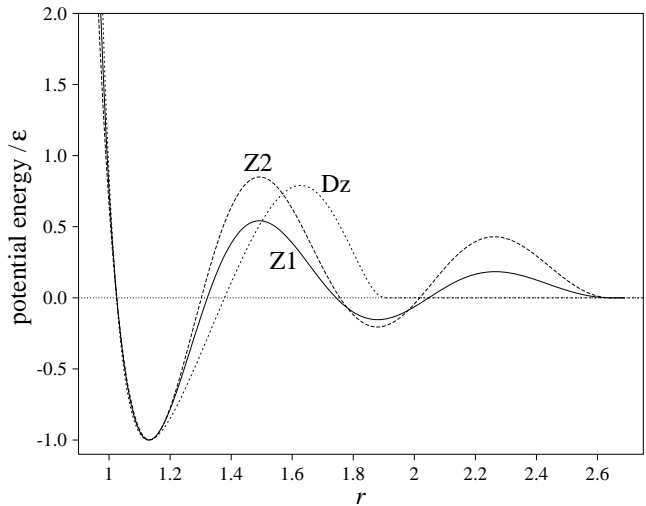}
\caption{\label{fig:potentials}The potential energy curves of the two new
potentials (Z1 and Z2) compared to the original Dzugutov (Dz) potential.
To aid comparison of the potentials, the energy is in units of $\epsilon$, 
the pair well depth for the appropriate potential.
}
\end{center}
\end{figure}

An alternative approach is to use a monatomic system, but one that has been 
tailored to prevent crystallization. There have been two notable ways in which this goal 
has been achieved. In the first, a small term is added to the potential (in this case
a Lennard-Jones potential) that is
a function of the static structure factor and inhibits ordering.\cite{DiLeonardo00} 
However, under some conditions this method is not completely effective at 
preventing crystallization.\cite{Angelani02}
The second approach involves designing a potential that disfavours the close-packed order 
that is usually favoured by monatomic systems with isotropic pair potentials.
Dzugutov achieved this by introducing a maximum in the potential at approximately
$\sqrt 2$ times the equilibrium pair distance (see Figure \ref{fig:potentials}),\cite{Dzugutov92,Dzugutov93b} 
which is the distance between opposite vertices of the octahedra that 
are intrinsic to close-packing. This maximum causes the preferred local order
in the system to be polytetrahedral and icosahedral.
Thus, systems bound by this Dzugutov potential
are good glass-formers, and exhibit a first sharp diffraction
peak and a split second peak in the structure factor,\cite{Sadigh99} which are common features
of metallic glasses. One other interesting feature of this potential is that
under certain conditions it can produce a dodecagonal quasicrystal.\cite{Dzugutov93}

It is well known that there is an increasing tendency for
local polytetrahedral and icosahedral order in simple liquids as the 
temperature is decreased.\cite{Jonsson,NelsonS,Yonezawa91} 
Recent results for Dzugutov liquids have shown how this change in local order 
affects the overall structure of the liquid, and is responsible for
the changes in properties as the glass transition is approached. 
There is low-dimensional growth of icosahedral clusters, which eventually percolate
through the system leading to a dramatic increase in the structural relaxation 
time.\cite{Zetterling01,Dzugutov02b} 
The described cluster growth leads to two interesting
dynamical effects: a pronounced spatial variation of atomic mobility
and a breakdown of the Stokes-Einstein relation.

An understanding of this intriguing behaviour can be obtained by examining
the structures of isolated clusters.\cite{Doye01a} 
The lowest-energy structures of the clusters can all be considered as 
aggregates of linked 13-atom icosahedra. However, these aggregates are not compact and as 
the size increases they change from chains to rings to porous three-dimensional network
structures. Therefore, the clusters show the same tendency for low-dimensional 
aggregation of icosahedra.

One additional advantage of using a monatomic system to study supercooled liquids
is that it is much easier to understand the effect of changing the form 
of the potential. We can therefore systematically study how
properties, such as the glass-forming ability and fragility, depend on
the potential. This is the approach we use here.

One justification for the form of the Dzugutov potential is that the
maximum resembles the first of the Friedel oscillations that occur for
metal potentials. However, further maxima occur in the
latter case. Here we investigate two potentials that have two maxima. 
The crystallization behaviour of one of these potentials (Z1) has already 
been studied by molecular dynamics simulations\cite{Dzugbrass}---the 
$\gamma$-brass\cite{Brandon74} structure resulted---and 
crystallization patterns of the other pair potential (Z2) 
will be explored elsewhere.\cite{DzugNaZn13} 
In this paper, we focus on the lowest-energy structures
exhibited by clusters bound by these potentials, paying particular
attention to the relationship between the observed structures and the
form of the potential.

Complementary to these studies of isolated clusters, we also explore the bulk 
liquids for the two potentials with molecular dynamics.
One particular purpose of these simulations is to investigate the liquids' behaviour
upon supercooling and their glass-forming abilities. We also compare
the lowest-energy structures for the isolated clusters with the
coherently structured domains with predominantly icosahedral local
order that develop upon supercooling of the bulk liquids. 
The results we present here will be
particularly helpful in rationalizing the role of the pair potential
and the related local order in the development of extended transient
structures in the supercooled liquids as the glass transition is
approached.

\section{Methods}

The two potentials both have the form 
\begin{equation}
V(r)=a {e^{\alpha r}\over r^3} \cos(2k_F r)+b\left({\sigma\over r}\right)^{n}+V_0 
\end{equation}
for $r<r_c$ and 0 otherwise.
We use the position of the third minimum in the function as our cutoff distance, $r_c$,
and $V_0$ is defined through the equation $V(r_c)=0$,
i.e.\ $V_0$ acts to shift the potential so that it vanishes at the third minimum, thus making the 
function and its first derivative continuous at the cutoff.
The values of the parameters for the two potentials are given in Table \ref{table:params},
and they are plotted in Fig.\ \ref{fig:potentials} together with the 
Dzugutov potential. We denote the two potentials as Z1 and Z2. 

The form of the new potentials has a more physical basis than the original Dzugutov potential. 
The first term has a form similar to that expected for the effective interaction between 
metal ions when screened by electrons. Friedel oscillations are present with wave-vector, 
$2k_F$, where $k_F$ corresponds to the wave-vector at the Fermi level. 
The second term adds a repulsive interaction that suppresses the oscillations at small $r$.
The potentials look similar to the effective pair potentials 
often derived for metallic systems,\cite{Hafner88b,Moriarty97} 
however in the latter case the potentials are density dependent.

\begin{table}
\caption{\label{table:params}Values of the parameters for the Z1 and Z2 potentials.
The values of $r_c$ and $V_0$ are defined by the relations in the text, and
only truncated values are reproduced here.}
\begin{ruledtabular}
\begin{tabular}{ccccccccc}
   &  $a$ & $\alpha$ & $k_F$ &       $b$        & $\sigma$ &  $n$ &    $r_c$ & $V_0$ \\
\hline
Z1 & 1.58 &    -0.22 & 4.120 & $4.2\times 10^8$ &    0.331 & 18.0 & 2.64909 & 0.04682632 \\
Z2 & 1.04 &     0.33 & 4.139 & $4.2\times 10^7$ &    0.348 & 14.5 & 2.64488 & 0.13391543 \\
\end{tabular}
\end{ruledtabular}
\end{table}

The total potential energy of a cluster interacting with this potential is simply 
$E=\sum_{i<j} V(r_{ij})$. 
A cluster structure that has a low energy must have 
pair distances that lie close to the minima in the potential and that avoid the maxima.
The first maximum occurs at $r=1.320\,r_{\rm eq}$ and $1.318\,r_{\rm eq}$ for Z1 and Z2 respectively,
where $r_{\rm eq}$ is the appropriate equilibrium pair separation.
The potential still has a large positive value at $\sqrt{2}\,r_{\rm eq}$, 
thus the possibility of close-packed structures is 
ruled out, as for the Dzugutov potential.
Instead the potential favours polytetrahedral structures that can avoid this maximum;
if the tetrahedra are regular, the first next-nearest neighbour distances occur at 
$r=\sqrt{2/3}\,r_{\rm eq}=1.633\,r_{\rm eq}$,
close to the second minimum at $r=1.663\,r_{\rm eq}$ (Z1) or $1.659\,r_{\rm eq}$ (Z2). 

\begin{figure}
\includegraphics[width=8.4cm]{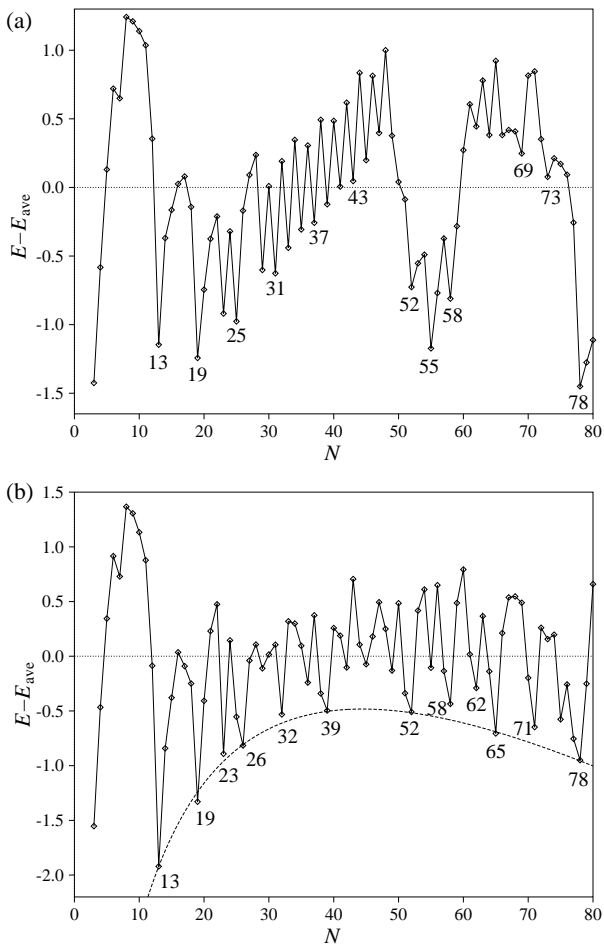}
\caption{\label{fig:Egmin}Energies of clusters interacting with the two potentials. 
In (a) the energy zero is $E_{\rm ave}^{\rm Z1}=-2.9445\,N + 0.9278\,N^{2/3} -2.7108\,N^{1/3} + 10.1508$
and in (b) is 
$E_{\rm ave}^{\rm Z2}= -2.7311\,N + 1.0818\,N^{2/3} - 7.4369\,N^{1/3} + 15.9735$. 
In (b) the dashed line represents the fit to the energies of the chains of
icosahedra occurring at $N=13n$, $E_{\rm n-icos}=-2.6613\,N - 0.5940\,N^{2/3} + 1.8063\,N^{1/3} + 0.6760 $.
}
\end{figure}

As a packing of regular tetrahedra is not achievable in Euclidean space, polytetrahedral
structures inevitably involve nearest-neighbour distances that deviate from the ideal value.
The energetic penalty for this strain depends sensitively on the width of the potential well. 
As the first minima of the Z1 and Z2 potentials are somewhat narrower 
than the Dzugutov potential well (Figure \ref{fig:potentials}), 
one expects the lowest-energy structures to be non-compact, as for the Dzugutov clusters.\cite{Doye01a} 
This allows the strain energy inherent to compact polytetrahedral forms to be avoided, whilst
maintaining polytetrahedral order.
This effect can be quantified by matching the curvatures of the potential at $r=r_{\rm eq}$
to that for the Morse potential, which has a single parameter $\rho$ that determines the width 
of the potential well. $\rho_{\rm eff}^{Z1}=8.63$ and $\rho_{\rm eff}^{Z2}=8.89$, which are
somewhat larger than the value for the Dzugutov potential ($\rho_{\rm eff}^{Dz}=7.52$)\cite{Doye01a}
and much larger than the values of $\rho$ for which the Morse potential has compact polytetrahedral
global minima.\cite{Doye97d}

We attempted to locate the cluster global minima using the basin-hopping
global optimization algorithm.\cite{WalesD97,Li87a} 
Since the Z1 and Z2 potentials vanish beyond a distance $r_c$, precautions were needed
to maintain efficiency. The cluster is placed in a spherical container to prevent atoms 
evaporating. However, for the structures considered in the present work there can
still be a significant amount of empty space in the container. We therefore 
checked periodically whether the pair energy of any atoms was identically zero.
Such atoms were reconnected with the cluster by moving them within the attractive range of their
nearest neighbour.

The global minima of these clusters are particularly
hard to locate because the oscillations in the potential lead to a rough energy landscape 
with huge numbers of minima and large energy barriers
between different structural forms.\cite{Doye01a} Furthermore, the non-compact nature
of the global minima means that the volume of configuration space that needs to be searched
is much larger than for compact clusters.
Due to these difficulties, it would be no surprise if some of the lowest-energy minima that
we found at larger sizes are not the true global minima.
Indeed, for some sizes we were able to construct
lower-energy structures using the physical principles deduced from the
structures that we were able to find. 
However, as we are more interested in the general structural evolution of the
global minima with size, rather than the specifics of individual cluster sizes, 
this is not a serious problem.

\begin{table*}
\caption{\label{table:gmin}Energies and point groups (PG) of the lowest-energy minima located
for clusters interacting with the Z1 and Z2 potentials.}
\begin{ruledtabular}
\begin{tabular}{ccccccccccccc}
 & \multicolumn{2}{c}{Z1} & & \multicolumn{2}{c}{Z2} & & & \multicolumn{2}{c}{Z1} & & \multicolumn{2}{c}{Z2} \\
\cline{2-3}\cline{5-6} \cline{9-10}\cline{12-13} 
$N$ & PG & Energy & & PG & Energy & & $N$ & PG & Energy & & PG & Energy \\
\hline
3  & $D_{3h}$    &   -2.087454 & &   $D_{3h}$   &   -2.248317 & &  42  & $C_s$      & -111.113917 & &   $C_s$   & -111.615125 \\
4  & $T_d$       &   -4.174907 & &   $T_d$      &   -4.496633 & &  43  & $D_{3h}$   & -114.526364 & &   $C_s$   & -113.534760 \\
5  & $D_{3h}$    &   -6.363630 & &   $D_{3h}$   &   -6.891923 & &  44  & $C_{3v}$   & -116.579062 & &   $C_1$   & -116.860294 \\
6  & $C_{2v}$    &   -8.658495 & &   $C_{2v}$   &   -9.439398 & &  45  & $C_s$      & -120.057909 & &   $C_s$   & -119.765131 \\
7  & $D_{5h}$    &  -11.602168 & &   $D_{5h}$   &  -12.682192 & &  46  & $C_s$      & -122.284996 & &   $C_2$   & -122.234952 \\
8  & $C_s$       &  -13.873626 & &   $C_s$      &  -15.054741 & &  47  & $C_1$      & -125.544082 & &   $C_1$   & -124.643068 \\
9  & $C_{2v}$    &  -16.763730 & &   $C_{2v}$   &  -18.089069 & &  48  & $C_1$      & -127.783337 & &   $C_1$   & -127.608371 \\
10  & $C_{3v}$   &  -19.689189 & &   $C_{3v}$   &  -21.205084 & &  49  & $C_s$      & -131.249095 & &   $C_1$   & -130.709014 \\
11  & $C_{2v}$   &  -22.642702 & &   $C_{2v}$   &  -24.379370 & &  50  & $D_{2h}$   & -134.429471 & &   $C_1$   & -132.812275 \\
12  & $C_{5v}$   &  -26.171418 & &   $C_{5v}$   &  -28.242064 & &  51  & $C_{2v}$   & -137.399683 & &   $C_1$   & -136.352523 \\
13  & $I_h$      &  -30.518777 & &   $I_h$      &  -32.957039 & &  52  & $C_{4h}$   & -140.882324 & &   $C_1$   & -139.240474 \\
14  & $C_{3v}$   &  -32.585210 & &   $C_{3v}$   &  -34.743529 & &  53  & $C_1$      & -143.552544 & &   $C_1$   & -141.027989 \\
15  & $C_{2v}$   &  -35.223551 & &   $C_{2v}$   &  -37.132647 & &  54  & $C_s$      & -146.333923 & &   $C_1$   & -143.549558 \\
16  & $C_s$      &  -37.875551 & &   $C_s$      &  -39.558764 & &  55  & $C_s$      & -149.860950 & &   $C_1$   & -146.978638 \\
17  & $C_s$      &  -40.662152 & &   $C_s$      &  -42.515739 & &  56  & $C_1$      & -152.301798 & &   $C_1$   & -148.936593 \\
18  & $C_s$      &  -43.724272 & &   $C_s$      &  -45.495502 & &  57  & $C_s$      & -154.749296 & &   $C_2$   & -152.431591 \\
19  & $D_{5h}$   &  -47.665071 & &   $D_{5h}$   &  -49.388337 & &  58  & $C_{2h}$   & -158.032574 & &   $C_1$   & -155.443175 \\
20  & $C_{2v}$   &  -50.006755 & &   $C_{2v}$   &  -51.271893 & &  59  & $C_s$      & -160.349915 & &   $C_1$   & -157.230650 \\
21  & $C_1$      &  -52.475722 & &   $C_s$      &  -53.433094 & &  60  & $C_1$      & -162.641511 & &   $C_1$   & -159.634564 \\
22  & $C_s$      &  -55.150884 & &   $C_s$      &  -55.980724 & &  61  & $C_1$      & -165.152433 & &   $C_1$   & -163.117336 \\
23  & $D_{3h}$   &  -58.697886 & &   $D_{3h}$   &  -60.133544 & &  62  & $C_1$      & -168.159905 & &   $C_1$   & -166.132881 \\
24  & $C_s$      &  -60.937135 & &   $C_{3v}$   &  -61.878219 & &  63  & $C_1$      & -170.671345 & &   $C_1$   & -168.183340 \\
25  & $D_{5d}$   &  -64.432022 & &   $D_{5d}$   &  -65.353500 & &  64  & $C_2$      & -173.914159 & &   $C_1$   & -171.393944 \\
26  & $C_s$      &  -66.463833 & &   $D_{2d}$   &  -68.387900 & &  65  & $C_1$      & -176.220446 & &   $C_1$   & -174.666612 \\
27  & $C_s$      &  -69.042269 & &   $C_2$      &  -70.379358 & &  66  & $C_2$      & -179.608322 & &   $C_1$   & -176.454127 \\
28  & $C_s$      &  -71.735736 & &   $C_{2v}$   &  -72.997963 & &  67  & $C_1$      & -182.418355 & &   $C_1$   & -178.831638 \\
29  & $C_s$      &  -75.412716 & &   $C_{2v}$   &  -75.975750 & &  68  & $C_1$      & -185.275450 & &   $C_1$   & -181.526788 \\
30  & $C_s$      &  -77.640330 & &   $C_1$      &  -78.606894 & &  69  & $C_s$      & -188.283757 & &   $C_1$   & -184.286813 \\
31  & $D_{5h}$   &  -81.114676 & &   $D_{5h}$   &  -81.271817 & &  70  & $C_1$      & -190.562830 & &   $C_1$   & -187.676283 \\
32  & $C_s$      &  -83.136786 & &   $C_s$      &  -84.659842 & &  71  & $C_1$      & -193.380298 & &   $C_1$   & -190.827932 \\
33  & $D_{3d}$   &  -86.608012 & &   $C_1$      &  -86.557138 & &  72  & $C_s$      & -196.722616 & &   $C_1$   & -192.621016 \\
34  & $C_{3v}$   &  -88.660767 & &   $C_s$      &  -89.323678 & &  73  & $C_s$      & -199.845668 & &   $C_1$   & -195.425747 \\
35  & $C_{2v}$   &  -92.153818 & &   $C_s$      &  -92.270480 & &  74  & $C_1$      & -202.558045 & &   $C_1$   & -198.085004 \\
36  & $C_s$      &  -94.381048 & &   $C_s$      &  -95.348047 & &  75  & $C_1$      & -205.448065 & &   $C_1$   & -201.559019 \\
37  & $D_{5d}$   &  -97.784539 & &   $C_s$      &  -97.471834 & &  76  & $C_i$      & -208.375341 & &   $C_1$   & -203.939219 \\
38  & $D_{6h}$   &  -99.874893 & &   $C_2$      & -100.923541 & &  77  & $C_s$      & -211.573361 & &   $C_1$   & -207.134890 \\
39  & $C_s$      & -103.330875 & &   $C_2$      & -103.814208 & &  78  & $C_{2v}$   & -215.615804 & &   $C_1$   & -210.028014 \\
40  & $C_s$      & -105.563799 & &   $C_1$      & -105.793804 & &  79  & $C_1$      & -218.290788 & &   $C_1$   & -212.027576 \\
41  & $C_s$      & -108.883830 & &   $C_{2v}$   & -108.594370 & &  80  & $C_2$      & -220.977578 & &   $C_1$   & -213.815040 \\
\end{tabular}
\end{ruledtabular}
\end{table*}

\section{Results}
\subsection{Isolated Clusters}

\begin{figure*}
\begin{center}
\includegraphics[width=16cm]{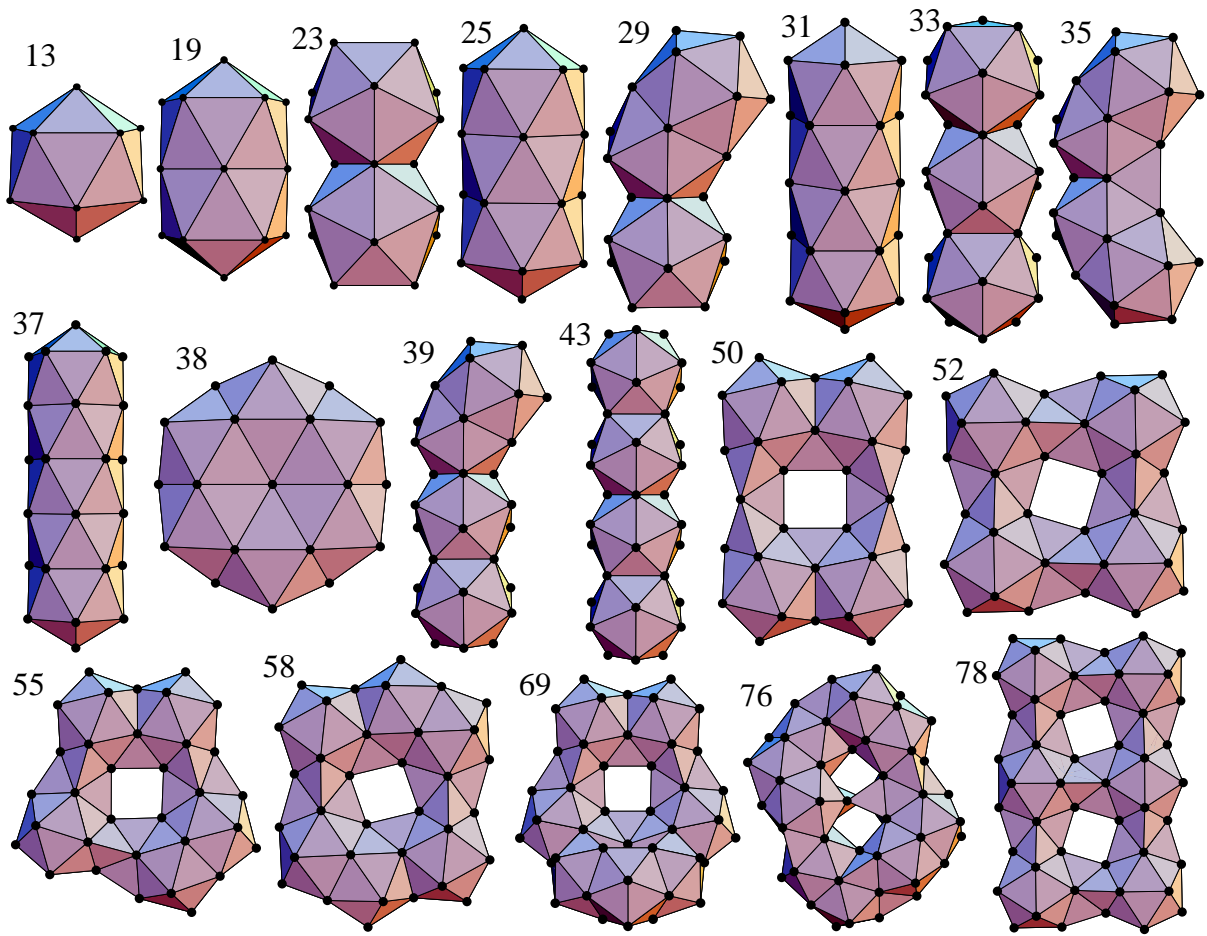}
\caption{\label{fig:pic_gamma}A selection of Z1 global minima. 
Each is labelled with the value of $N$.}
\end{center}
\end{figure*}

\begin{figure*}
\begin{center}
\includegraphics[width=16cm]{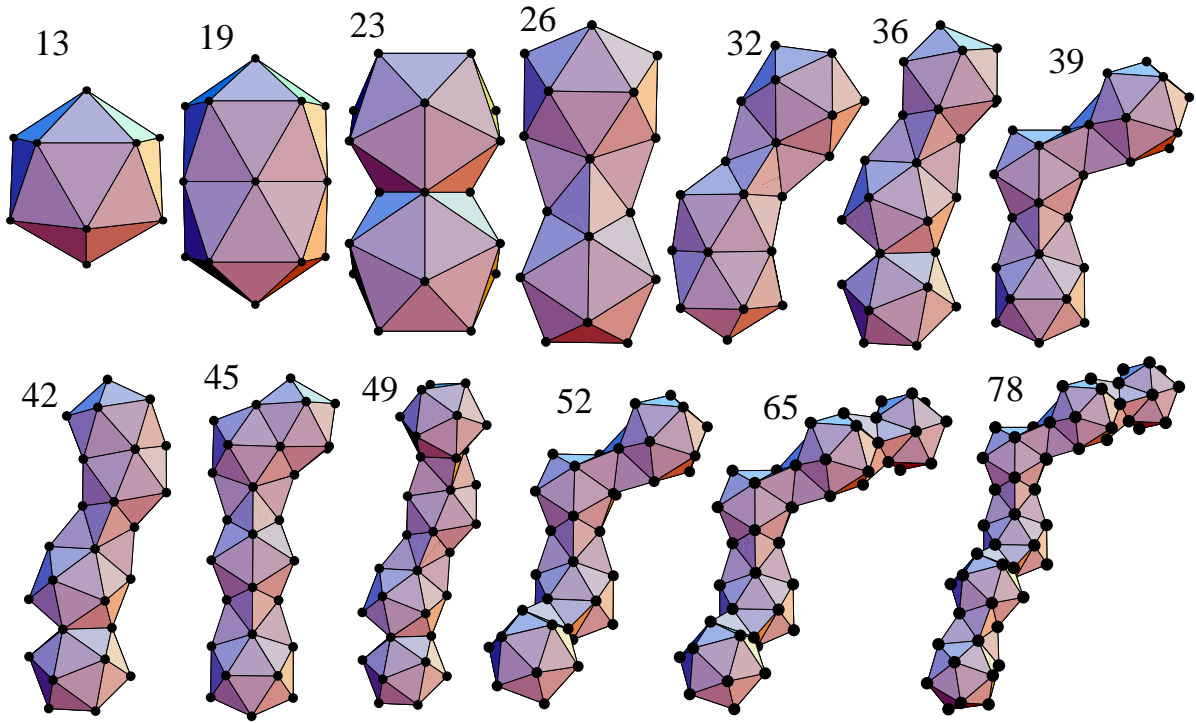}
\caption{\label{fig:pic_nxtal}A selection of Z2 global minima. Each is labelled with the value of $N$.}
\end{center}
\end{figure*}

The energies and point groups of the putative global minima for 
all clusters up to $N=80$ that we have located for the 
two potentials are given in Table \ref{table:gmin}. 
The points files will also be made available online at 
the Cambridge Cluster Database.\cite{Web} In Figure \ref{fig:Egmin} we plot the 
energies of the global minima in a way that causes particularly stable sizes to stand out.
The structures of a selection of these ``magic'' sizes, and other sizes where interesting
high symmetry structures were observed, are depicted in Figures \ref{fig:pic_gamma}
and \ref{fig:pic_nxtal}.

As expected, for both potentials the structures exhibited are non-compact arrangements 
of connected 13-atom icosahedra. For all the particularly stable sizes evident in 
Figure \ref{fig:Egmin} these icosahedra are complete.

There are three ways in which the icosahedra are linked: (i) two icosahedra can 
interpenetrate sharing a common fivefold axis, as in the 19-atom double icosahedron; 
(ii) two icosahedra can share a face as in the 23-atom structure that is the 
global minimum for both potentials; (iii) two separated icosahedra can be joined by 
a tetrahedron whose opposite edges are shared with the two icosahedra, as in the 26-atom
structure that is the Z2 global minimum. All these ways of linking icosahedra
are common in metallic alloys.\cite{Shoemaker}

These three different modes of linkage 
leads to a large number of possible icosahedral aggregates, especially at larger sizes.
For example, the 55-atom Z1 global minimum involves all three possibilities: there is one pair
of interpenetrating icosahedra, one pair of face-sharing icosahedra and 
three pairs of icosahedra that are bridged by tetrahedra.

For both potentials, beyond $N=13$ the global minima initially develop into chains 
of linked icosahedra. 
For the Z1 potential these chains involve a mixture of both interpenetrating and face-sharing 
icosahedra, but no icosahedra linked by tetrahedra. Such structures with complete icosahedra
are only possible at odd sizes, leading to the odd-even oscillations seen in 
Figure \ref{fig:Egmin}(a) for $N=29-49$. The only global minimum in this size range that is not 
chain-like occurs at $N=38$. It is a disk-like fragment of the Z-phase (one of the 
Frank-Kasper phases\cite{FrankK58,FrankK59}), which was particularly stable for the 
Dzugutov potential.\cite{Doye01a}  Beyond $N=50$ the Z1 global minima start to exhibit ring-like structures, 
and at the largest sizes that we consider they become double rings. 

The 52-atom Z1 global minimum is particularly interesting. It is a square array of 
icosahedra linked through tetrahedra, and it is easy to see how this motif can be repeated
to form extended two- and three-dimensional networks of icosahedra. 
For example, the 78-atom double ring structure is simply formed by 
the addition of two further icosahedra. 
Furthermore, two of the 52-atom structures can be placed
on top of each other to produce a cubic array of icosahedra at $N=104$. 
Although we have not attempted any global optimization at this size, a comparison
of the energy of this cluster ($-296.947505$) with an extrapolation of $E_{\rm ave}$ 
(a fit to the energies of the putative global minima) suggests that it is extremely stable.
This result has two implications. Firstly, the transition to three-dimensional structures is likely
to occur for the Z1 potential just beyond the size range we consider. Indeed, the 91-atom 
structure formed by removal of an icosahedron at one corner of the cube also appears 
to be particularly stable ($E=-253.432268$).
Secondly, it seems plausible that at zero pressure the crystal structure formed by 
repeating this cubic unit could also be most stable. This crystal would be the 
same as the NaZn$_{13}$ crystal\cite{Shoemaker} but with vacancies at the Na sites. 
This possibility contrasts somewhat with the results of constant volume simulations 
(corresponding to higher densities) where crystallization 
to a $\gamma$-brass structure\cite{Brandon74} was observed.\cite{Dzugbrass} 

For the Z2 potential, by contrast, there is a definite preference for icosahedra linked by 
tetrahedra. Hence, there is a series of stable chain structures that occur at $N=13n$ (Fig.\ \ref{fig:Egmin}). 
Furthermore, these are the most stable structures over the whole size range that we study. 
There is no crossover to two- or three-dimensional networks of icosahedra.
When we add a third icosahedron to the 26-atom global minimum, there are five possible 
positions for it. In fact neither the linear nor the L-shaped configurations are lowest in energy.
Instead, the angle between the central atoms of the three icosahedra in
the 39-atom global minimum is approximately 120$^\circ$. This angle might lead one to think
that a hexagonal ring could be formed from six icosahedra, but because the linked icosahedra 
are in different orientations, a flat configuration with four icosahedra is not possible. 
Instead, the 52-atom global minimum is a bent chain with two 120$^\circ$ angles 
between the centres, and a 90$^\circ$ torsion angle.
It is also noteworthy that the extra linkage between icosahedra in the ring of 
the Z1 52-atom global minimum is not enough to stabilize this structure,
because the Z2 potential so disfavours 90$^\circ$ angles between the icosahedral centres. 
However, in constant volume bulk simulations for the Z2 potential, the system
has been observed to crystallize into a NaZn$_{13}$-like structure with most of the 
Na sites vacant.\cite{DzugNaZn13}

When comparing the behaviour of the structures for the Z1 and Z2 potentials with the Dzugutov clusters, it
is clear that the Z1 clusters are the most similar to the Dzugutov clusters. 
However, the crossovers to two- and to three-dimensional
icosahedral aggregates occur at larger sizes for the Z1 potential,
and the Dzugutov potential has a greater preference for face-sharing linkages.\cite{Doye01a}
Therefore, there is an increasing tendency for the clusters 
to exhibit non-compact, low-dimensional icosahedral aggregates 
on going from the Dzugutov, to the Z1 and then to the Z2 potential. 
One can say that these systems become more frustrated,\cite{Sadoc99} i.e.\ it 
becomes increasingly hard to propagate the locally preferred polytetrahedral order 
freely through space. There is also a modified version of the Dzugutov 
potential that has been designed to exhibit compact polytetrahedral clusters by giving the potential a wider well 
that can sustain the inherent strain in these clusters.\cite{Doye01d} 
The effective range parameter for this potential, $\rho^{\rm m-Dz}_{\rm eff}$, is 5.16.

The above series of four potentials therefore provides an ideal opportunity to correlate the properties 
of the supercooled liquid with the local structure seen in the clusters and the form of the 
potential. As the potential becomes more frustrated, one would expect
a percolating network of icosahedral clusters to develop more rapidly as the temperature
is decreased, thus leading to more fragile behaviour,\cite{Angell95} 
i.e.\ the structural relaxation time has a super-Arrhenius temperature dependence. 
Similarly, rearrangements of these icosahedral aggregates are also likely to be especially slow, and thus they
inhibit crystallization. So one also expects the more frustrated potentials to be better glass-formers.

\begin{figure}
\begin{center}
\includegraphics[width=8.5cm]{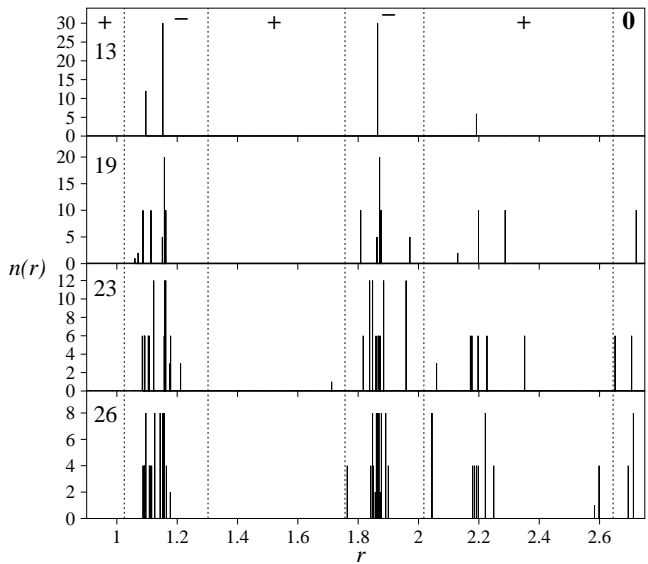}
\caption{\label{fig:dist}
The number of pairs separated by a distance $r$, $n(r)$, for the Z2 global minima at 
$N$=13, 19, 23 and 26, as labelled. The dotted lines correspond to the distances at 
which the Z2 potentials changes sign or goes to zero at the cutoff distance.}
\end{center}
\end{figure}

It is relatively easy to relate the differences between the structures observed for 
the different types of cluster back to the form of the potential. 
First, we shall examine the variations in the tendency to form low-dimensional aggregates.
The second maximum in the Z1 and Z2 potentials places an additional constraint on the system, as
it is harder to avoid than the first maximum. Distances at roughly $2 r_{\rm eq}$ in polytetrahedral
clusters (e.g.\ opposite vertices of an icosahedron) occur near the centre of this maximum (Fig.\ \ref{fig:dist})
and so are significantly penalized. The number of such distances is of course larger for a compact 
compact cluster, so the Z1 and Z2 potentials have an extra force driving them
towards non-compact geometries over and above that for the Dzugutov potential. 
This effect is more pronounced for the Z2 clusters, simply because of the larger
size of the maximum (Figure \ref{fig:potentials}). 

It is also possible to rationalize the preferred types of linkage between the icosahedra.
To aid this analysis we have decomposed the potential into five terms
\begin{equation}
\label{eq:Edecomp}
E=-n_{\rm nn} \epsilon +E_{\rm strain} + E_{\rm max1} + E_{\rm min2} + E_{\rm max2},
\end{equation}
where $n_{\rm nn}$ is the number of nearest neighbours, $E_{\rm strain}$ is the energetic 
penalty for nearest-neighbour distances that deviate from the equilibrium pair separation,\cite{Doye97d} and 
$E_{\rm max1}$, $E_{\rm min2}$ and $E_{\rm max2}$ are the energies associated with pair 
distances near to the first maximum, second minimum and second maximum, respectively. 
The distance criteria used to separate these different contributions are simply the 
distances for which $V(r)=0$. 

We report the above contributions to the energy
for the icosahedron and the three ways of linking two icosahedra in Table \ref{table:Edecomp}.
We also show the pair distribution functions for these structures in Fig.\ \ref{fig:dist}.
For a given size, the number of nearest neighbours is largest for the 
interpenetrating icosahedra and smallest for the icosahedra linked by tetrahedra, 
as would be expected from the relative compactness of the resulting structures.\cite{nnchain}
However, to compensate for this effect the interpenetrating icosahedra have the largest strain energy
and $E_{\rm max2}$ for a given size and the tetrahedra-bridged icosahedra the smallest. 

When comparing the contributions to the energy for the same structure, but for the two potentials,
it is clear from Table \ref{table:Edecomp} that the major changes are in $E_{\rm min2}$ and 
$E_{\rm max2}$, due to the greater magnitude of the second maximum and minimum for the Z2 potential.
Indeed, there is a greater driving force for minimizing $E_{\rm max2}$ for the Z2 potential, and
it is this factor that leads to the icosahedra linked by an intervening tetrahedron to 
be dominant.

Similarly, when linking more than two icosahedra through intervening tetrahedra, 
90$^\circ$ angles between the icosahedral centres are disfavoured for the Z2 potential because 
this introduces some distances between the atoms in the non-adjacent icosahedra that 
are near to the second maximum. 

\begin{table}
\caption{\label{table:Edecomp}
The contributions to the energy defined by Equation (\ref{eq:Edecomp}) for the 13-atom icosahedron
and the three ways of linking them.
All the energies are measured in units of $\epsilon$, the pair well depth for the appropriate potential.
}
\begin{ruledtabular}
\begin{tabular}{cccccccc}
& $N$ & Energy & $n_{\rm nn}$ & $E_{\rm strain}$ & $E_{\rm max1}$ & $E_{\rm min2}$ & $E_{\rm max2}$ \\
\hline
Z1 & 13 & -43.860 & 42 & 1.711 & 0.000 &  -4.531 & 0.959 \\
   & 19 & -68.502 & 68 & 4.258 & 0.000 &  -8.442 & 3.682 \\
   & 23 & -84.358 & 84 & 5.429 & 0.088 & -10.841 & 4.965 \\
   & 26 & -94.894 & 92 & 3.779 & 0.000 & -11.586 & 4.913 \\
\hline
Z2 & 13 & -43.976 & 42 & 1.787 & 0.000 &  -6.075 &  2.312 \\
   & 19 & -65.900 & 68 & 4.454 & 0.000 & -11.096 &  8.742 \\
   & 23 & -80.238 & 84 & 6.151 & 0.162 & -14.282 & 11.730 \\
   & 26 & -91.252 & 92 & 4.315 & 0.000 & -15.322 & 11.755 \\
\end{tabular}
\end{ruledtabular}
\end{table}

\subsection{Bulk liquids}

The properties of the bulk liquids for the Z1 and Z2 potentials 
were explored by molecular dynamics simulations of
a system of $16\,000$ particles at a number density $\rho$=0.84. 
The simulations focused on the behaviour of these liquids in the 
supercooled domain. We take as the upper bound of this domain, not the limit
of thermodynamic stability (the melting temperature $T_m$), but the
temperature ($T_A$) below which the liquid begins to demonstrate
a set of characteristic dynamical anomalies that are routinely
associated with the notion of supercooled liquid dynamics. The most
celebrated of these anomalies is the super-Arrhenius slowing down,
which is commonly regarded as a defining property of the
supercooled liquid state. One purpose of the present molecular dynamics
simulations was to explore the glass-forming abilities of the two
systems (the notion of the glass-forming ability of a liquid refers to
its ability to remain in a state of metastable equilibrium within a
temperature domain where it exhibits the super-Arrhenius behaviour and
other characteristic features of supercooled liquids). To this
end, we subjected each of the two liquids to a step-wise cooling
whereby at each temperature step the system was relaxed to the state of 
apparent equilibrium. 

\begin{figure}
\includegraphics[width=8.4cm]{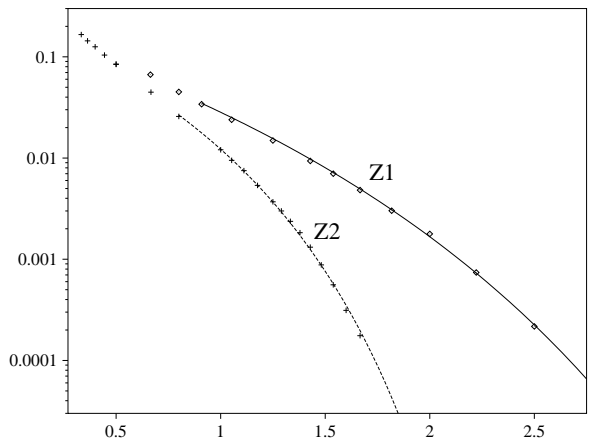}
\caption{\label{fig:Arr} An Arrhenius plot showing the temperature dependence of the
diffusion constant for bulk Z1 and Z2 liquids, as labelled. 
The data points are simulation results and the solid lines are fits to the
data using the VTF (Vogel-Tammann-Fulcher) equation, $D=D_0 \exp[BT_0/(T-T_0)]$.
}
\end{figure}

The temperature variation of the diffusion coefficient for both
potentials is shown in Fig.6. In both cases, $D(T)$ exhibits a pattern
characteristic of fragile glass-formers: below the 
temperature $T_A$ a pronounced super-Arrhenius slowing down
is observed. The non-Arrhenius part of the $D(T)$ curves extends for
over two decades, thus demonstrating the pronounced glass-forming
ability of the two systems simulated here. We also note that the
activation energy, as estimated from the Arrhenius plot, is significantly
larger in the case of the Z2 potential; this behaviour appears to be
consistent with the larger amplitude of its Friedel-like oscillations.

Another issue that we addressed in these simulations was the formation 
of extended domains of icosahedral structure upon supercooling. The
development of such domains has previously been demonstrated in 
simulations of a supercooled bulk liquid using
the Dzugutov potential.\cite{Zetterling01,Dzugutov02b} Moreover, these
domains were found to be morphologically similar to the low-energy structures
of isolated clusters for that potential.\cite{Doye01a} 

Here, we analyse atomic configurations obtained by steepest descent
minimization of instantaneous atomic configurations representing
equilibrium liquids. In order to discern icosahedral order in the
first shell of neighbours, we first identified pentagonal bipyramids---pairs 
of neighbours with five common neighbours forming a closed ring
(any two particles separated by the distance less than 1.5 were
regarded as neighbours). A 13-atom icosahedron was identified as a
central atom with 12 neighbours all of which form five-fold pairs
(pentagonal bipyramids) with the central atom. Furthermore, we
considered two icosahedra sharing at least 3 atoms as connected.  

\begin{figure}
\includegraphics[width=8cm]{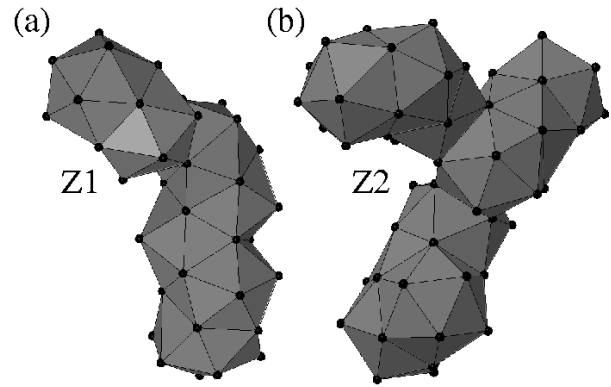}
\caption{\label{fig:liqclust} 
Clusters of connected icosahedra that were located in the 
bulk liquids at $T$=1.0 and number density $\rho$=0.84. 
(a) a 53-atom cluster from the Z1 liquid and 
(b) a 73-atom cluster from the Z2 liquid.
}
\end{figure}

We found that both systems also demonstrate a 
clear tendency for the formation of extended icosahedral
clusters upon cooling. This tendency is illustrated in Fig.\ 
\ref{fig:liqclust} where we present typical clusters of connected 
icosahedra that develop upon supercooling of the two model liquids. 
For both systems, it is possible to observe a close similarity between 
the patterns of icosahedral aggregation in these clusters and those 
in the respective clusters presented in Figs. \ref{fig:pic_gamma} 
and \ref{fig:pic_nxtal}. 
We remark that the Z2 system clearly demonstrates, for both
the isolated clusters and the icosahedral domains identified
in the liquid, a more pronounced tendency for one-dimensional
aggregation of icosahedra than the Z1 system.

\section{Conclusions}

Following Frank,\cite{Frank52} we have sought insight into the structure
of supercooled liquids by analysing the structures of the corresponding clusters;
the clusters can allow the identification of the ideal, local structural order.
This philosophy has also been employed in cluster models of the glass transition,\cite{Fan02}
and has influenced theories such as the frustration-limited domain model.\cite{Kivelson95,Kivelson98}

We have introduced two new potentials that are similar to
effective potentials in metals with oscillations that lead to 
polytetrahedral and icosahedral order being preferred, thus making them
good glass-formers. The cluster global minima generally correspond 
to non-compact arrangements of linked 13-atom icosahedra. 
For one potential (Z1) we find that two-dimensional networks of these
icosahedra become preferred for the largest sizes we consider, whereas
for the other potential (Z2) chains of icosahedra are only observed. 
Similar icosahedral aggregates were observed to form upon supercooling
of the corresponding bulk liquids.

Along with the Dzugutov potential\cite{Dzugutov92,Dzugutov93b} 
and a modified version of it,\cite{Doye01d} this series of potentials
provides an ideal set of model systems to explore the relationship between the
potential, the locally preferred structure, the energy landscape and the 
properties of supercooled liquids, such as fragility, glass-forming ability 
and the decoupling of diffusion and structural relaxation.
Understanding these relationships has been a long-standing goal in the field 
of supercooled liquids, and simulations, here and in more detail elsewhere,\cite{DzugNaZn13}   confirm 
some of the expected correlations with the cluster structures we observe. 

\begin{acknowledgments}
JPKD is grateful to the Royal Society for financial support.  FHMZ and
MD gratefully acknowledge support from the Swedish Research Foundation
(VR).
\end{acknowledgments}


\begin{thebibliography}{43}
\bibitem[{nnc()}]{nnchain}
\bibinfo{note}{This result can be proved by calculating the number of nearest
  neighbours in a chain of icosahedra as a function of $N$ when linked in the
  different ways. $n_{\rm nn}^i=26 N/6-43/3$, $n_{\rm nn}^f=42 N/10-63/5$,
  $n_{\rm nn}^t=50 N/13-8$ where $i$, $f$ and $t$ stand for interpenetrating,
  face-sharing and tetrahedra bridged, respectively. For $N>13$ $n_{\rm nn}^i >
  n_{\rm nn}^f > n_{\rm nn}^t$.}
\end{thebibliography}
\end{document}